\documentstyle{mn}
\input psfig.tex

\title[Near-IR spectroscopy of CSS sources]{Near-infrared spectroscopy
of powerful compact steep-spectrum radio sources}
\author[P.~Hirst~et~al.]{Paul~Hirst,$^{1,2,3}$\thanks{Email
p.hirst@jach.hawaii.edu}
Neal~Jackson,$^1$
Steve~Rawlings,$^4$\\
$^1$University of Manchester, Jodrell Bank Observatory, Macclesfield, Cheshire. 
SK11 9DL.\\
$^2$X-ray Astronomy Group, Dept. Physics \& Astronomy, University of Leicester, 
LE1 7RH. \\
$^3$Joint Astronomy Centre, 660 N. A'ohoku Place, Hilo, HI 96720, USA\\
$^4$Astrophysics, NAPL, Keble Road, Oxford. OX1 3RH}

\date{\today.}

\pagerange{\pageref{firstpage}--\pageref{lastpage}}
\pubyear{1997}

\newcommand{\othree}{[O{\sc iii}]}
\newcommand{\fetwo}{Fe{\sc ii}}
\newcommand{\otwo}{[O{\sc ii}]}
\newcommand{\stwo}{[S{\sc ii}]}
\newcommand{\ntwo}{[N{\sc ii}]}
\newcommand{\halpha}{H$\alpha$}
\newcommand{\hbeta}{H$\beta$}

\begin{document}
\label{firstpage}

\maketitle

\begin{abstract}

\noindent
We have obtained near-infrared spectroscopy of a small sample of powerful Compact
Steep-Spectrum (CSS) radio sources mainly, but not exclusively, from the
3CR sample. We find no differences between the distributions in the
equivalent width and luminosity of the \othree 5007\ line for our 
sample and other larger, presumably older, 
high-redshift 3C objects, suggesting that the underlying quasar
luminosity remains roughly constant as quasars age. We also find a
possible broad line in 3C~241, adding to recent evidence for broad lines
in some radio galaxies.
\end{abstract}



\section{INTRODUCTION}

Compact Steep-Spectrum (CSS) radio sources are an important subclass of
radio sources, which form around 15 -- 30 per cent of radio source
surveys (e.g. Peacock \& Wall 1982; Fanti et al. 1990; O'Dea 1998). 
They are characterised by a small projected linear size $D$ (typically
$<$15~kpc) and a steep high-frequency radio spectrum ($\alpha<-0.5$
where $S_{\nu}\propto\nu^{\alpha}$).

Fanti et al. (1990) presented an important catalogue of CSS sources 
still in use by many authors, by selecting from the earlier catalogues
of Peacock \& Wall (1982) and the 3CR catalogue of Jenkins, Pooley \& Riley
(1977). They analysed the statistics of CSS sources compared
to normal, larger radio sources and concluded that there are too many
CSS sources observed in complete samples to be consistent with the
hypothesis that they are normal sources seen in projection.
Neither are they likely to be an intrinsically faint, and hence numerous,
population of normal radio sources
seen in projection and amplified by Doppler boosting into the sample. 
This result follows because the majority of their radio luminosity
originates from twin radio lobes, which cannot be Doppler boosted as
such an effect would cause large and unobserved asymmetries between
the brightnesses of the two lobes. It is therefore likely that
CSS sources are a class of source whose difference from larger sources
needs to be explained by physical or evolutionary models. A
comprehensive review (O'Dea 1998) summarises their properties and the
state of observational knowledge about these objects.


CSS sources may be small either because they are young (Phillips \&
Mutel 1982; Carvalho 1985) or because they exist in a dense environment 
which prevents
propagation of the radio jets (van Breugel et al. 1984). Large bends in
CSS radio jets suggest interaction with such an environment (e.g. Fanti
et al., 1986, 1989). However, there are several observations which do
not favour denser environments in CSS than normal radio sources. 
ISO far-infrared fluxes of CSS objects are broadly consistent with 
those of larger radio sources (Fanti et al. 2000) suggesting similar
dust content in the circumnuclear environment. Near-infrared nuclear 
continuum properties are also similar (de Vries et al. 1998) and optical
spectroscopy reveals little difference in [OII] fluxes for a given
radio flux in low-redshift CSS and extended objects (Morganti et al. 1997).
Moreover, direct VLBI observations of compact symmetric objects
(CSOs) which have structure on the sub-kiloparsec scale, reveal young
dynamical ages (Owsianik \& Conway 1998) and imply a natural evolutionary 
connection between CSO and CSS sources. Such connections
have been modelled by  number of authors (Bicknell, Dopita \& O'Dea 1997;
O'Dea \& Baum 1997; Snellen et al. 2000) who deduce the evolution of 
luminosity and radio source size with source age along the evolutionary 
sequence. The end result of CSO and CSS evolution is not clear; O'Dea
\& Baum suggest that such sources may eventually evolve into low-power
(Fanaroff \& Riley class I) sources rather than edge-brightened, 
powerful ``classical-double'' radio sources. It is also possible that in
addition there may exist a population of small sources which never reach
the stage of classical-double radio sources; Alexander (2000) uses such
an extra population to fit the density of sources in the power--linear
size plane.


Emission-line properties of CSS sources have 
been the subject of a number of optical
studies, usually involving spectroscopy or narrow-band imaging
both with ground-based telescopes and the Hubble Space Telescope.
Imaging reveals that CSS have much of their optical emission 
(between 30\% and 90\%) within the central 3 kiloparsecs,the
same dimensions as the typical radio source (Axon et al. 2000). In
addition, faint extended line emission was seen beyond the radio source in
four out of five cases for which sufficiently deep observations were 
available (Axon et al. 2000).
It seems likely that at least some of the emission near the centre is associated
with the radio jets, as there are many examples where the optical
line emission is either cospatial with the radio jets or at
least elongated in the same direction (de Vries et al. 1999;
Axon et al. 2000), or where complex moderately 
broad ($\sim 500$~km$\,$s$^{-1}$) line profiles (Gelderman \& Whittle
1994) are seen close to the radio lobes (O'Dea et al. 2002; Morganti
et al. 1997) suggesting that the radio lobes are exciting the
line-emitting gas by driving shocks into it. 
The fainter line emission beyond the radio sources (Axon et al. 2000) 
may be produced by photoionization by anisotropically
escaping photons emitted by a central active galactic nucleus. The twin
roles of radiative shocks and accretion disks as sources of
photoionizing photons have been discussed recently by Inskip et al.
(2002).

%

In large, non-CSS radio sources, much work has been done on the
luminosities of narrow optical emission lines, in particular \otwo
3727\AA\ and \othree 5007\AA\ . There appears to be a distinction
between radio galaxies and quasars in that the \othree\ line may
be stronger for quasars than radio galaxies of the same radio
luminosity (Jackson et al. 1990).
This could be due to orientation-dependent obscuration
extending into the narrow-line region; models of such effects
have been made by Hes, Barthel \& Fosbury (1996), di Serego Alighieri et
al. (1997) and Baker \& Hunstead (1995).
On the other hand, \otwo\ luminosities of radio galaxies
appear to be similar to those of quasars (Hes et al. 1996). At higher
redshift, where the \othree\ line is shifted into the near infra-red,
Jackson \& Rawlings (1997) failed to find a significant difference in
its luminosity between radio galaxies and quasars. One can speculate
that this may be because the greater radio power of these high-redshift
objects' jets leads directly to the removal of obscuration of \othree\ 
(Lawrence 1991). Alternatively there may be subtle reasons, as suggested
by Simpson (1998), why one expects emission-line luminosity differences 
between radio galaxies and quasars which are dependent on the line studied
and the ranges of redshift and radio luminosity probed.
Based on simple photoionization models, Simpson (1998) also suggests that
``\othree\ is a better indicator than \otwo\ of the strength of the 
underlying [quasar] continuum''.

If there is a connection between CSS and larger sources, either
because CSS are destined to evolve into larger objects or because
they contain a denser environment which prevents them so evolving,
we may be able to learn about this connection by studying the
narrow lines which are produced within the host galaxy of the CSS
which also completely contains the radio source. Baker \& Hunstead
(1995) found that the optical continuum of CSS quasars was 
reddened compared to that of larger sources and that the equivalent
widths of narrow emission lines was correspondingly high, although they
were not able clearly to disentangle suppression of the continuum from
enhancement of the narrow lines. Morganti et al.
(1997) have studied this problem for low-redshift objects and
conclude that \othree\ and \otwo\ luminosities are similar to those of
larger sources. 

This paper is a continuation of earlier work (Jackson \& Rawlings 1997) 
in which we carried out infra-red spectroscopy of the \othree\ emission
line in larger 3CR radio sources and found that this
emission line in radio galaxies and quasars has
approximately the same strength for a given redshift.  
Here we extend the programme to CSS sources. Our aims are (i) to test
whether the Morganti et al. (1997) result (similar emission-line
luminosities in CSS and larger sources) also holds at high redshift, and
(ii) to investigate whether CSS radio galaxies and quasars at high
redshift have similar emission-line luminosities. The results have
implications for the physical conditions in high-redshift CSS sources,
and have implications for evolutionary models in which
CSS sources eventually expand to become large radio sources. A
subsidiary aim is to investigate whether any of the objects hitherto
classified as CSS galaxies possess infra-red broad lines and should
therefore be reclassified as quasars; we have therefore taken spectra of
the region around \halpha\ where this is possible. For the main part of
the work we use \othree\ as it is generally stronger than
\otwo\ and also because there are suggestions (Best,
R\"ottgering \& Longair 2000; Inskip et al. 2002) 
that \otwo\ luminosities may be more affected by shock excitation. In
section 2 we describe the sample, observations and data reduction; in
section 3 we present the results and in section 4 discuss the relation
between our results and other previous work. Section 5 contains a
brief discussion of the implications of these results.

\section{SAMPLE SELECTION, OBSERVATIONS AND DATA REDUCTION}

\subsection{Sample selection}

Fanti et al. (1990) selected a complete sample of objects from the 3CR
(Jenkins et al. 1977) and Peacock \& Wall (1982) samples with the
following selection criteria:

\begin{enumerate}
\item Linear size of 15 kpc or less, assuming $H_0$=100 km s$^{-1}$ Mpc$^{-1}$
and $q_0$=1
\item 178-MHz radio power, $P_{178}$, of $10^{26.5}$ WHz$^{-1}$ or
greater
\item 178-MHz flux density of 10~Jy or greater. Since the radio spectra of CSS
sources often turn over at frequencies of a few hundred MHz, sources
were included if the extrapolation of their high-frequency radio spectra
exceeded 10~Jy at 178~MHz. 
\item High-frequency radio spectral index
$\alpha<-$0.5 [where $S_{\nu}\propto\nu^{\alpha}$, and following Peacock
\& Wall (1982) the high-frequency spectral index is taken between 2.7~GHz
and 5~GHz by most subsequent studies].
\end{enumerate}

Table~\ref{tab:sample} gives the basic characteristics of the objects
which were observed. In all, seven CSS quasars and two CSS radio
galaxies were observed.

Our original sample contained all seven CSSs from Fanti
et al. (1990) with RAs in the range 05:30 -- 15:00 with $z>1$, 
declination $\delta<60^{\circ}$ for accessibility to UKIRT 
and with \othree\ lines away from any strong atmospheric sky line. Six
of these were observed, the exception being 3C~287. Of these, two 
(OQ172 and 4C31.38) were included, following Fanti et al. (1990),
 because the extrapolation of their 
high-frequency radio spectra exceeded 10~Jy at 178~MHz although their
actual $S_{\rm 178MHz}<10$~Jy. Three slightly lower redshift sources, 
3C~147, 3C~237 and 3C~286 from Fanti et al. (1990) were also observed, 
picked entirely because of RA accessibility. Note that we assume 
$H_0$=50 km s$^{-1}$ Mpc$^{-1}$ and $q_0$=0 throughout, which gives 
linear sizes a factor of 3 greater than Fanti et al. (1990) for a redshift
1 object. Redshifts in the table are from the compilation of Spinrad et al.
(1985) except for 4C~31.38 from Fanti et al. (1990) and OQ~172 from 
Morton et al. (1989). $R$ values (the fraction of radio flux from the
core of the source, see Orr \& Browne 1982) are at an emitted frequency
of 8GHz, assuming radio spectral indices, $\alpha$ of 0 and -1 for the nuclear 
and extended emission, from Saikia et al. (1995). We do not quote an $R$
value for OQ~172 as it is difficult to determine the radio morphology and
core identification from radio maps of this source in the literature
(e.g. Udomprasert et al. 1997). OQ~172 is a gigahertz peaked spectrum
(GPS) source (e.g. O'Dea, Baum \& Stanghellini 1991). 
Linear sizes are taken from the compilation of Fanti et al. (1990) which
in turn are derived mostly from MERLIN and VLA maps of Spencer et al.
(1989) at wavelengths of 18cm, 6cm and 2cm, and
correcting to our assumed cosmology. Such maps have typical resolutions 
of between 0\farcs2 and 0\farcs5, which is sufficient to derive linear
sizes and core fluxes unambiguously in the large majority of cases. 
178-MHz radio flux densities
are taken from Kellerman, Pauliny-Toth \& Williams (1969) except for
1153+31 (Pilkington \& Scott 1965) and 1442+10 (K\"uhr et al. 1981; only
the 365-MHz flux was available).

\begin{table*}
\caption{Overview of the sample. We give the name and IAU designation of
each object, together with the type (Quasar or Galaxy), the value of
$R$, defined as the radio of core radio flux density to extended flux
density at 8 GHz in the emitted frame, the angular size $D$ (see text)
and the radio power at 178 MHz (except for OQ172 for which the 365-MHz
power is quoted and which is marked with a dagger). We give the linear 
size for both ($H_0$, $q_0$) =
(50,0.0) as used in this paper 
and (100,1.0) as in the definition of Fanti et al. (1990).}
\label{tab:sample}
\begin{tabular}{llcllccc}
\hline
Object  & IAU name & Type  & $z$   & $R$  & $D$ & $D$ & log$_{10}(P_{178})$\\
        & (B1950) &        &       &      & /kpc & /kpc &/log$_{10}$WHz$^{-1}$)\\
        &         &        &       &      &(50,0)&(100,1)& \\
\hline
3C 147  & 0538+498      & Q     & 0.545 & 0.15   & 5.9 & 2.4 & 29.10 \\
3C 186  & 0740+380      & Q     & 1.063 & 0.052  & 24  & 8.2 & 29.20 \\
3C 190  & 0758+143      & Q     & 1.197 & 0.093  & 46  & 14.1 & 29.37 \\
3C 237  & 1005+077      & G     & 0.877 & 0.00097& 12  & 4.5 & 29.15 \\
3C 241  & 1019+222      & G     & 1.617 & 0.0063 & 15  & 2.8 & 29.63 \\
4C 31.38& 1153+317      & Q     & 1.56  & 0.005  & 11  & 3.1 & 29.39 \\
3C 286  & 1328+307      & Q     & 0.849 & 0.82   & 39  & 14.2 & 29.18 \\
3C 298  & 1416+067      & Q     & 1.439 & 0.094  & 31  & 9.1 & 30.09 \\
OQ 172  & 1442+101      & Q     & 3.544 &        & $<$1& $<$0.4 & 30.01$\dagger$ \\
\hline
\end{tabular}
\end{table*}

\subsection{Observations}

The observations were carried out using the short focal length camera
in the CGS4 spectrometer
(Mountain et al. 1990) on the United Kingdom Infrared telescope (UKIRT) on 
Mauna Kea, Hawaii. At the time, the spectrograph used a
58$\times$62-pixel detector array. 
Flux and wavelength calibrated $J$, $H$ and $K$ band spectra 
were obtained of the \othree\ and \halpha\ regions of a sample of 9 objects 
at medium to high redshifts. Observations in J-band were taken using a
75~lines$\,$mm$^{-1}$ grating in second order, giving wavelength
coverage of 0.17~$\mu$m. Two grating settings were used in order to give
wavelength coverage of 1.00-1.17~$\mu$m or 1.16-1.34~$\mu$m, according to
the redshift of the object. In $H$-band a 150~lines$\,$mm$^{-1}$
grating was used in second order, giving a wavelength coverage of 
0.09~$\mu$m and in $K$-band a 75~lines$\,$mm$^{-1}$ grating was used in
first order, giving wavelength coverage of 0.4~$\mu$m. 
A slit projecting to one pixel at the detector (3\farcs 1
on the sky) was used in all cases. This led to a resolving power of 
about 400 in J-band, 500 in H-band and 300 in K-band
(about 800, 600 and 1000~km$\,$s$^{-1}$ respectively). 
In the observations, the detector was
stepped in a four-point, half-pixel step-size 
pattern in wavelength in order to provide good 
sampling of the spectral resolution and redundancy over bad pixels. 
The detector was also stepped 10
pixels across the sky on alternate integrations in an ABBAABBA pattern.

Observations were carried out over a period of 3 nights beginning on
1994 March 15. Conditions were mainly spectrophotometric, although 
thin cirrus was observed during the observation of OQ172.
The radio galaxy
observations were all made using blind offsets as they were too faint to
see on the telescope monitor system.

\begin{table}
\caption{Parameters of the observations, including the source name,
exposure time, spectral band and wavelength, grating used, diffraction
order, and spectral resolution. 
The exposure time is per spectral point in each case; the
total time spent on each source is eight times the exposure time per
point.}
\label{tab:obs}
\begin{tabular}{lcccc}
\hline
Source  & Exp.  &Band   &Grating& Res.  \\
	& time &      &/Order   & (nm) \\
        & /s  & /l$\,$mm$^{-1}$  &  \\
	\hline
3C 147  & 360   & J (1.00--1.18$\mu$m)    & 75    / 2     & 2.8   \\
3C 186  & 240   & J (1.00--1.18$\mu$m)    & 75    / 2     & 2.8   \\
3C 190  & 320   & J (1.00--1.18$\mu$m)    & 75    / 2     & 2.8   \\
3C 237  & 360   & J (1.16--1.35$\mu$m)    & 75    / 2     & 2.8   \\
3C 241  & 320   & J (1.16--1.35$\mu$m)    & 75    / 2     & 2.8   \\
3C 241  & 240   & H (1.68--1.78$\mu$m)    & 150   / 2     & 3.5   \\
4C 31.38& 400   & J (1.16--1.35$\mu$m)    & 75    / 2     & 3.3   \\
3C 286  & 160   & J (1.16--1.35$\mu$m)    & 75    / 2     & 3.3   \\
3C 298  & 120   & H (1.56--1.66$\mu$m)    & 150   / 2     & 3.3   \\
OQ 172  & 2530  & K (2.0--2.4$\mu$m)    & 75    / 1     & 6.5   \\
\hline
\end{tabular}
\end{table}

\subsection{Data reduction}

Data reduction followed the standard procedures described by 
Eales \& Rawlings (1993). Bias subtraction and flat fielding 
were carried out by the online telescope data reduction system. Subsequent
steps were carried out using the Image Reduction and Analysis
Facility (IRAF) package distributed by the National Optical Astronomy
Observatory (NOAO), and analysis of the one-dimensional spectra was
performed using software written by one of us (NJ). 
Wavelength calibration was performed using argon arc lines, or
atmospheric OH lines in the case where an arc spectrum was not
available. The lines were identified using the IRAF {\sc identify} task,
and a set of 2-D transformation parameters were generated using the IRAF
{\sc reidentify} task and applied using IRAF {\sc transform} to correct
for geometric distortion across the chip. 
Background sky lines were subtracted from the corrected frames
by subtracting the B spectra from the A spectra in the ABBAABBA chopping
pattern described above. The resulting difference frames contained a 
``negative'' channel containing the object as seen in the B exposures 
and a ``positive'' channel from the A exposures; 
the spectra were then extracted. They 
were flux calibrated using observations of stars from
the UKIRT list of standard stars, which were assumed to emit black body
spectra of a temperature determined from their published spectral type.
Error frames representing the standard deviation on each pixel were also 
propagated through the reduction process, the expected standard deviation
being determined using photon count statistics, assuming a Poisson 
distribution for the number of photons arriving at each pixel. Flux
calibration errors can be estimated by the dispersion of the flux
calibration curves for different standard stars observed during the
programme, and are approximately 20\%.

The luminosities given in Table~\ref{tab:lines} have been calculated from
fluxes taken by directly integrating the flux under the line in the
spectra. Continua were fitted to the data by eye, and the adopted
continuum fit in each case is shown on the right-hand panel of each
figure. Errors on the line fluxes are of the order of 10--20\%
based on photon statistics, together with a larger systematic error
due to continuum fitting. In typical cases, moving the continuum within
a plausible range gives changes at the 10-20\% level in the flux of the
\othree\ 500.7-nm line and at the $\sim$30\% level in the weaker lines of
\othree\ 495.9 nm and \hbeta . We
note, however, that the \othree\ lines at 495.9~nm and 500.7~nm have
been fitted separately, and in all cases the ratio of fluxes in these
lines is close to the 1:3 ratio which is expected theoretically. For the
two galaxies, an additional error may be introduced due to the blind
offsetting, although the large slit width (3\arcsec) compared to the
likely $\sim$0\farcs5 accuracy of the offset stars implies that this
error is relatively small.
In the case of good signal-to-noise spectra (3C~147, 3C~286,
3C~298) the widths of the emitted lines have been deduced from
the width of Gaussian fits to the observed lines using $W_{em} =
\sqrt{W_{obs}^2 - W_{psf}^2}$, $W_{psf}$ being the width of the
spectral point spread function of the spectrometer, taken from the
quoted spectral resolution of the CGS4 configuration in use, and checked
by comparison with the argon arc lines used for wavelength calibration,
which have full widths at half-maximum (FWHM) of 800$\pm$50~km$\,$s$^{-1}$.

In the case of spectra with poor signal-to-noise the fitting routine was
also used to estimate the parameters of each line (centroid, FWHM
and intensity) and the emitted width of the line,
$W_{em}$, was inferred as above. In each case, 1000 artificial spectra
were then made which had the same 
signal-to-noise and line strength as each individual spectrum 
being measured but using different random 
number seeds; this procedure was repeated for different intrinsic 
FWHM in order to assess the likelihood of each different intrinsic FWHM
reproducing our measurement. For conservatism,  a $W_{psf}$ 20\% greater
than that inferred from the CGS4 manual was used in the simulations. As
an example, we show in Fig. 1 the result of simulations for an
intrinsically very narrow line and different S:N ratios. We 
quote in Table 3 the probability that
the line we have measured is consistent with zero, based on these
simulations. We also quote the upper and lower 1-$\sigma$ bounds on the
intrinsic FWHM, although the error distribution is not Gaussian on the
intrinsic widths because this quantity is calculated as 
$\sqrt{W_{obs}^2 - W_{psf}^2}$ where $W_{obs}$ does have a Gaussian
error distribution. We discuss the interpretation of the narrow-line
widths further in section 4.1.

\begin{figure}
\psfig{figure=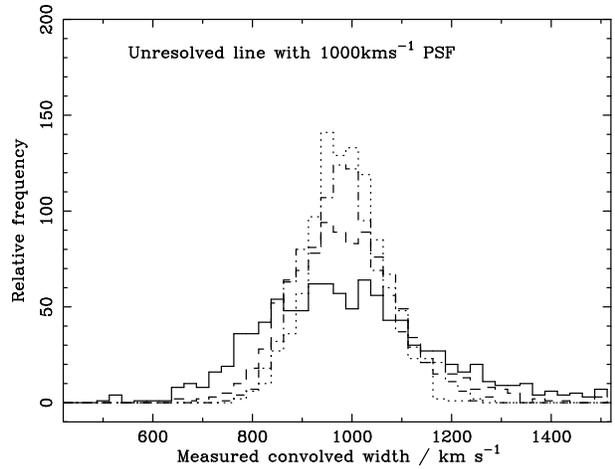,width=8cm,angle=-90}
\caption{Probability distribution for measured FWHM of spectral lines,
for an unresolved line convolved with a 1000~km$\,$s$^{-1}$ spectral
PSF. The spectra have been simulated for the same pixel spacing as the
actual J-band spectra (see text).
Solid, dashed, dash-dotted and dotted lines are for signal-to-noise
values of 7, 10, 13 and 16 per pixel in the peak of the line. Most of
the measured spectra have S:N of between 10 and 13 per pixel, allowing
detections of intrinsic 
line widths of about 700~km$\,$s$^{-1}$ (which result in
measured widths of 1200~km$\,$s$^{-1}$).}
\end{figure}

\begin{table}
\caption{Line luminosities, deconvolved (see text) full widths at half
maximum in km s$^{-1}$ and rest-frame equivalent widths of
observed lines (6717+6731\AA\ lines together in the case of \stwo ).
Widths marked with an asterisk represent lines that have a greater than
20\% chance of being less than the 800-km$\,$s$^{-1}$ resolution 
based on simulations. Luminosities of strong lines have errors between
20--50\% (see text) and those marked with an asterisk are determined
from line fluxes which are more than 20\% uncertain based on photon count
statistics. Lines which are marginally detected are marked ``det?''.}
\label{tab:lines}
\begin{tabular}{|l|c|c|c|c|}
\hline
Object  & Line          & log $L$ & FWHM  & EW \\
       &            &    /W   & /km$\,$s$^{-1}$ & /nm\\
\hline
3C 147  & \halpha +\ntwo     & 36.90 & 2570     & \\
        &                    &       & $\pm$150 & \\
        & \stwo        & 35.84 &          & \\
\ \\
3C 186  & \othree 5007 & 37.03 & 1100             & 4.6 \\
        &              &       & $\pm$400 (0.1\%) & $\pm$2.0 \\
        & \othree 4959 & 36.57 & *                & \\
        & \hbeta       & 36.68 & 3500             & \\
        &              &       &$\pm$1800         & \\
\ \\
3C 190  & \othree 5007     & 36.96 & 950            & 6.9 \\
        &                  &       & $\pm$400 (1\%) & $\pm$3.0\\
        & \othree 4959     & 36.49 &  *             & \\
        & \hbeta           & 36.04 & 2100           & \\
        &                  &       & $\pm$1000      & \\
\ \\
3C 237  & \halpha\ + \ntwo  & 36.71 & 2000    & \\
        &                   &       & $\pm$500 & \\
        & \stwo             & 36.26 &          & \\
\ \\
3C 241  & \othree 4959 & 36.67  & *                & \\
        & \othree 5007 & 37.08  &    1000          & 4.9 \\
        &              &        & $\pm$400 (3\%)   &$\pm$2.0 \\
        & \hbeta       & 36.76* & 1200             & \\
        &              &        & $\pm$700 (1\%)   & \\
        & \halpha      & 37.16  & 2000             & \\
        &              &        & $\pm$900 (0.5\%) & \\
        & \ntwo{6583}  &  det?  &               *  & \\
\ \\
4C~31.38& \othree 5007 & 36.46  & *           & 0.67 \\
        & \othree 4959 & det?  &              & \\
\ \\
3C 286  & \halpha\ + \ntwo& 37.52 & 2600      & \\
        &                 &       & $\pm$120  & \\
        &                 & det? &              & \\
\ \\
3C 298  & \halpha\ + \ntwo  & 38.27 & 3500    & \\
        &              &         & $\pm$170   & \\
\ \\
OQ~172  & \othree 5007 & 38.00 &   2200       & 5.0 \\
        &              &         & $\pm$600   &$\pm$2.0\\
        & \othree 4959 & 37.56* & *           & \\
        & \hbeta           & 37.93 & 3700     & \\
        &              &         & $\pm$1000  & \\
\hline
\end{tabular}
\end{table}

We have no definite evidence for extended emission from these objects;
spectra can be extracted from pixel rows adjacent to the object rows, which
show more flux than expected from the spill-over as determined from the
standard star observations, though these are better explained as
poor-positioning of the slit; the calibration stars are obviously much
brighter and thus easier to centre the slit on. 

\section{RESULTS AND NOTES ON INDIVIDUAL OBJECTS}

Fig. 2 shows the spectra obtained.
The parameters for identified lines are given in Table~\ref{tab:lines}.


\begin{figure*}
\begin{tabular}{c}
\psfig{figure=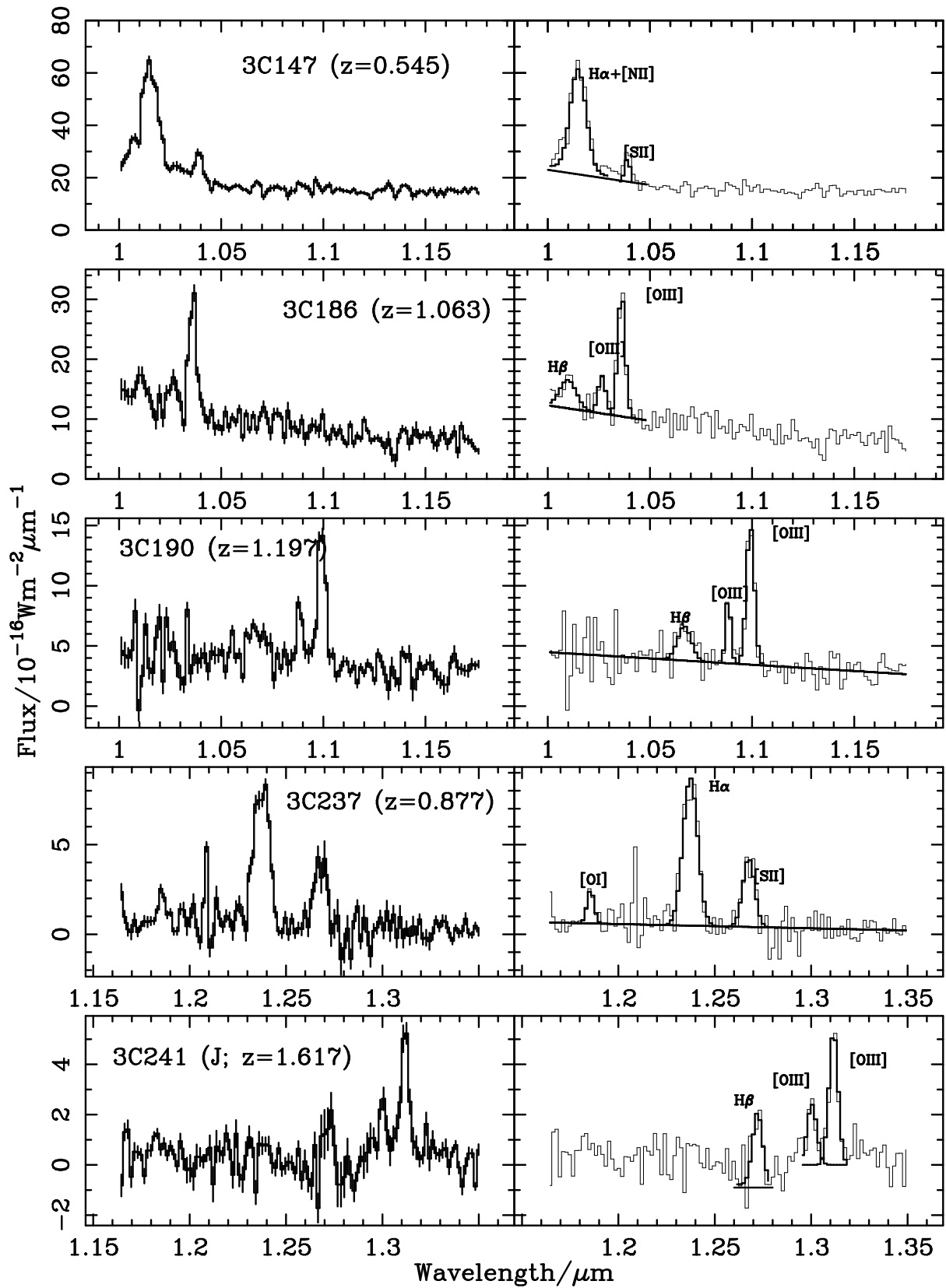,width=16cm}
\end{tabular}
\caption{The infra-red spectra. Each panel shows the spectrum obtained
on the left, including the error bars per pixel. The right-hand panel in
each case shows the adopted continuum level and line fits.}
\end{figure*}
\setcounter{figure}{1}
\begin{figure*}
\psfig{figure=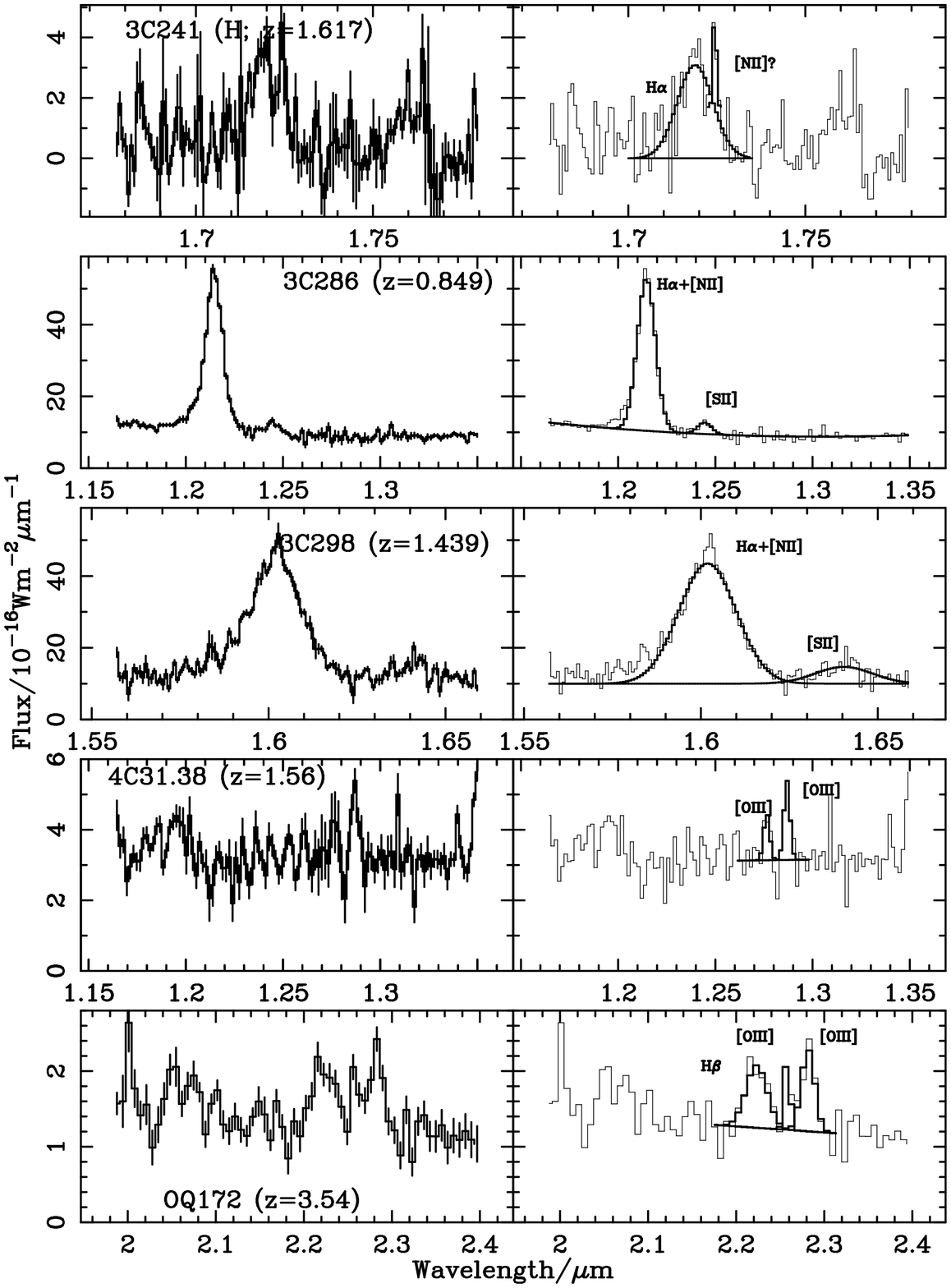,width=16cm}
\caption{The infra-red spectra, continued}
\end{figure*}

\begin{figure}
\psfig{figure=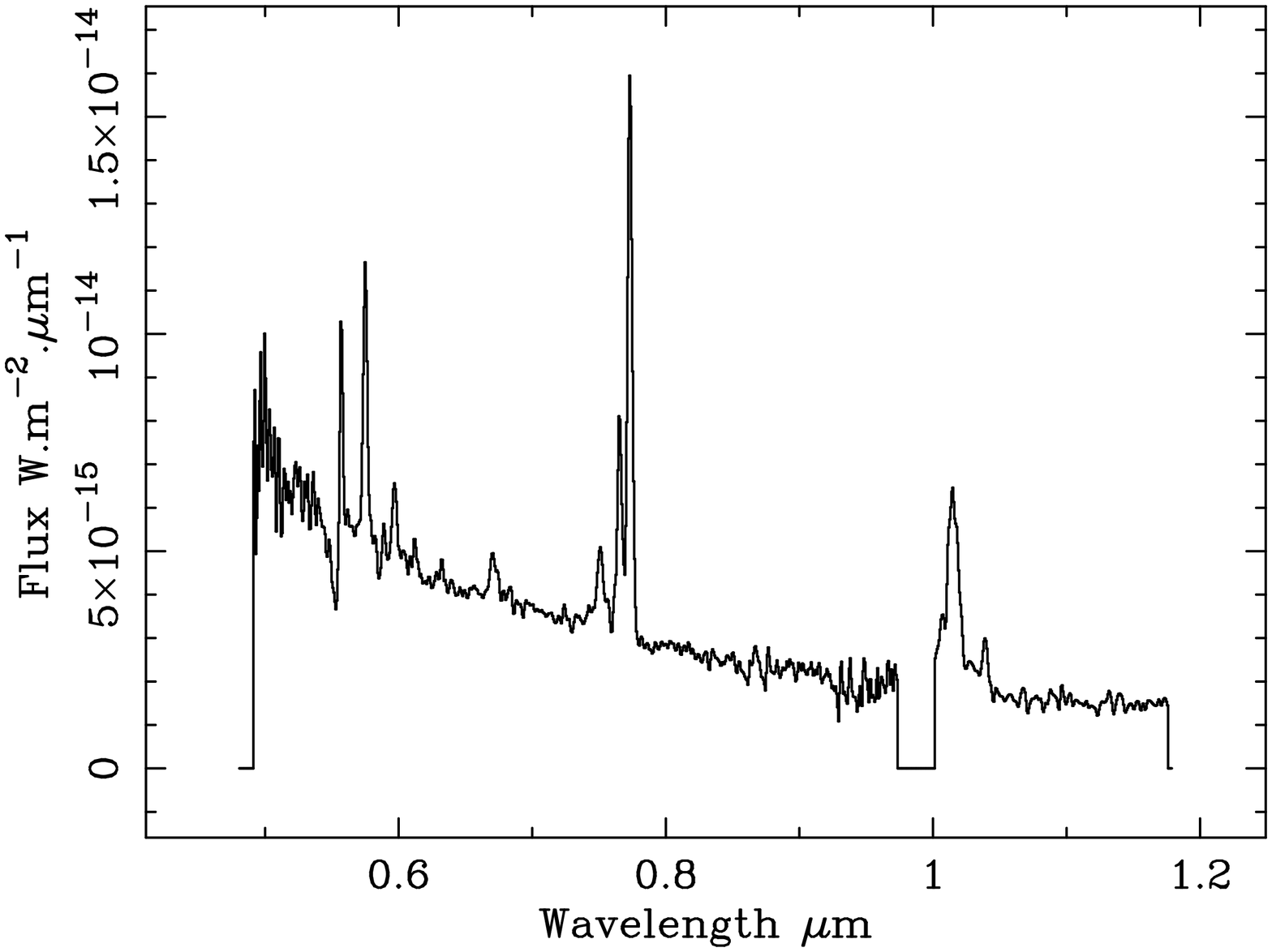,width=7cm,angle=-90}
\caption{Combined optical and infrared spectrum of 3C~147 from this paper and
Jackson \& Browne (1991)}
\label{fig:3c147comp}
\end{figure}

\begin{description}

\item[{\bf 3C~147}]
It is unclear if the very broad feature
surrounding the \halpha\ line is real, and if so, what its origin is. It is
possible that it is simply an artefact generated by failure of the flux
calibration process near the band edges; on the other hand, it may be a
broad wing of H$\alpha$. The error on the total \halpha\ flux is larger
than usual (probably 30-40\%) due to difficulty in fitting to the
continuum at the blue end of the spectrum. The noise
to the red end of the optical part of the composite is due to coincidence 
with a set of strong sky emission lines.
We can join our 3C~147 spectrum to the optical spectrum of Jackson \&
Browne (1991)
giving us confidence in the flux calibration of our IR data - the
power law optical continuum matches up in both spectral index and magnitude
between the two spectra. 
The composite spectrum thus obtained is shown in
Figure~\ref{fig:3c147comp}. 

\item[{\bf 3C~186}]
Although the spectrum is rather noisy, the continuum and \othree\ lines
are clearly detected, along with H$\beta$ at the blue end of the
spectrum. Continuum fitting is difficult in this object because of the
lack of continuum to the blue of \hbeta .

\item[{\bf 3C~190}]
The poor signal-to-noise ratio of the broad \hbeta\ line in our 3C~190 spectrum
makes it difficult to determine an accurate flux for this line.

\item[{\bf 3C~237}]
In our 3C~237 spectrum, it is doubtful whether the \stwo\ doublet lines 
have really been resolved - the signal-to-noise ratio is not good, and 
the doublet separation would be expected to be about equal to the minimum 
resolvable by the spectrometer. In any case, the errors on the pixels 
concerned make it impossible to determine any useful results from the 
relative fluxes of the doublet components. The spike near 1.2$\mu$m
appears as a single-pixel bright spot in the negative channel, and is
likely to be a cosmic ray.

\item[{\bf 3C~241}]

We have both J- and H-band spectra of 3C~241. In the J band spectrum, the
noticeable `dip' surrounding the base of the \hbeta\ line is due to 
unfortunate co-incidence with a broad atmospheric absorption line and
may also be due to poor sky subtraction. This 
prevents us from determining accurately the Balmer decrement for this 
source and the error on the flux is greater than usual (probably around
50\%). Each line in the \stwo\ doublet is marginally resolved.
Even though 3C~241 is classified as a narrow-line radio galaxy,
our H-band spectrum shows continuum and 
the possibility of broad \halpha\ emission.
In the following scatter diagrams, 3C~241 continues to be
marked as a radio galaxy, and is readily identifiable as 
the only high-redshift CSS radio galaxy for which we have 
\othree\ line data.

\item[{\bf 4C~31.38}]
The spectrum is noisy, but there is a detection of \othree\ 5007\AA\ and
a probable detection of \othree\ 4959\AA . The continuum is relatively
strong. 

\item[{\bf 3C 286}]
H$\alpha$ is clearly detected, but the \ntwo\ lines at 6548\AA\ and
6584\AA\ are too weak to be seen clearly. The \stwo\ 6717, 6731\AA\
lines are detected but not clearly resolved.

\item[{\bf 3C298}]
The H$\alpha$ line is clearly detected, and \ntwo\ is just visible on
the blue wing. There is some indication of the presence of \stwo\ lines
but they are not securely detected. The H$\alpha$ line may well contain
a narrow component which accounts for the failure to fit all the flux in
the centre of the line (e.g. Jackson \& Eracleous 1995). Alternatively
the broad line may have a profile which is more peaked than a Gaussian;
a Lorentzian profile gives a good fit to the profile including the
central region.

\item[{\bf OQ~172}]

Barthel et al. (1990) present an optical spectrum of OQ~172, which shows a
continuum flux consistent with extrapolation of our data.
The width of the \othree\ 5007\AA\ line in OQ172 is
severely affected by blending with the red wing of the strong H$\beta$
line and the width may be much lower than that quoted. 
\end{description}

\section{COMPARISON WITH OTHER RESULTS}

Our aim is to study the strengths of the \othree\ emission lines and the
properties of the Balmer lines in the CSS objects and to
compare them with non-CSS objects at similar redshift, for which spectra
were presented in Jackson \& Rawlings (1997) and with other
CSS objects at lower redshift and radio luminosity.

\subsection{Line profiles}

\subsubsection{Broad lines}

We note that the broad permitted lines in our spectra (\halpha\ and
\hbeta ) tend to have quite narrow widths, as compared to non-CSS
quasars at lower redshift -- an effect noted previously 
by Baker \& Hunstead (1996). It is not easy to find a comparison sample 
of non-CSS objects which is of comparable radio power and redshift; 
the only really comparable sample are 3CR quasars with $z>1$ for
which there are not many spectra yielding measurements of Balmer line
widths. Baker et al. (1994) took spectra of six high-redshift quasars, three
radio-quiet and three radio-loud, whose Balmer line widths are not
significantly different from those of our quasars, although the
comparison is complicated by the very much lower extended radio fluxes
of the Baker et al. quasars and by the fact that Baker et al. find that
their emission lines decompose naturally into a broad and a narrow
component\footnote{Formally, a Kolmogorov-Smirnov test is unable to 
distinguish our sample of FWHMs from those of Baker et al. (1994) 
at the 10\% significance level if we take the mean of Baker et al.'s 
broad and narrow component widths as the comparison FWHM, or if we just
take the narrow component widths.}.
Lacking ideal comparison data, we compare with samples at lower 
redshift and radio luminosity; the Molonglo Quasar Survey (MQS) for which
line-width measurements are presented by Baker et al. (1999) and the
compilation of line-widths collated by Wills \& Browne (1986). In the
former case, we exclude objects listed as CSS by Kapahi et al. (1998)
and in the latter case we use only the quasars whose radio structure is 
dominated by radio lobes in order to be consistent with the 3CR (and
predominant MQS) selection. 

A Kolmogorov-Smirnov test using our three measurements in quasars of 
H$\beta$ widths and our three measurements of
H$\alpha$ widths yields probabilities of 5\% and 4\% of being drawn
from the same distribution as those of Baker et al. (1999) and an 
11\% significance compared to Wills \& Browne's (1986) \hbeta\
measurements. Although it is not yet clear why this effect occurs,
Baker \& Hunstead (1996) suggest it could either be due to partial
reddening or intrinsically narrower broad-line-region lines. In
principle, \ntwo\ contamination could affect this comparison, although
this is unlikely because of the relative weakness of the \ntwo\ line and
the fact that narrow lines are may be visible in the spectra (e.g.
3C~298) and do not affect the broad line fit.

%
%

In 3C~241 the H$\alpha$ line may be broad. We discuss the implications 
of this and similar observations further in section 5.

\subsubsection{Narrow lines}

Although the simulations suggest that the \othree\ lines are broader
than would be expected from the theoretical resolutions and the widths
of the arc lines, a combination of continuum fitting problems and sky
subtraction residuals mean that it is difficult to make firm statements
about the \othree\ line widths and whether some of them are marginally
resolved; all are consistent at the 1$\sigma$ level with standard 
500--600~km$\,$s$^{-1}$ intrinsic line widths. This is also true of the study 
of Jackson \& Rawlings (1997) of larger sources which was performed 
with similar resolution and in which only one object (3C~191) clearly 
had \othree\ broader than a few hundred km$\,$s$^{-1}$. At
lower redshift, where high signal-to-noise observations are easily
available, Gelderman \& Whittle (1994) note ``broad structured \othree 
5007\AA\ line profiles'' in optical spectra of a sample of 19 lower-redshift
CSS objects. 

It is also possible that contaminating blends of \fetwo\ could be
responsible for broad wings to the \othree\ lines (see e.g. the spectra
of radio-quiet quasars analysed by Kuraszkiewicz et
al. 2000). In the case of radio-loud, and particularly radio-loud, 
lobe-dominated quasars, however, the \fetwo\
strength tends to be smaller than in radio-quiet quasars (e.g. Miley \&
Miller 1979; Heckman 1980).

\subsection{\othree\ properties versus radio luminosity}

\begin{figure}
\psfig{figure=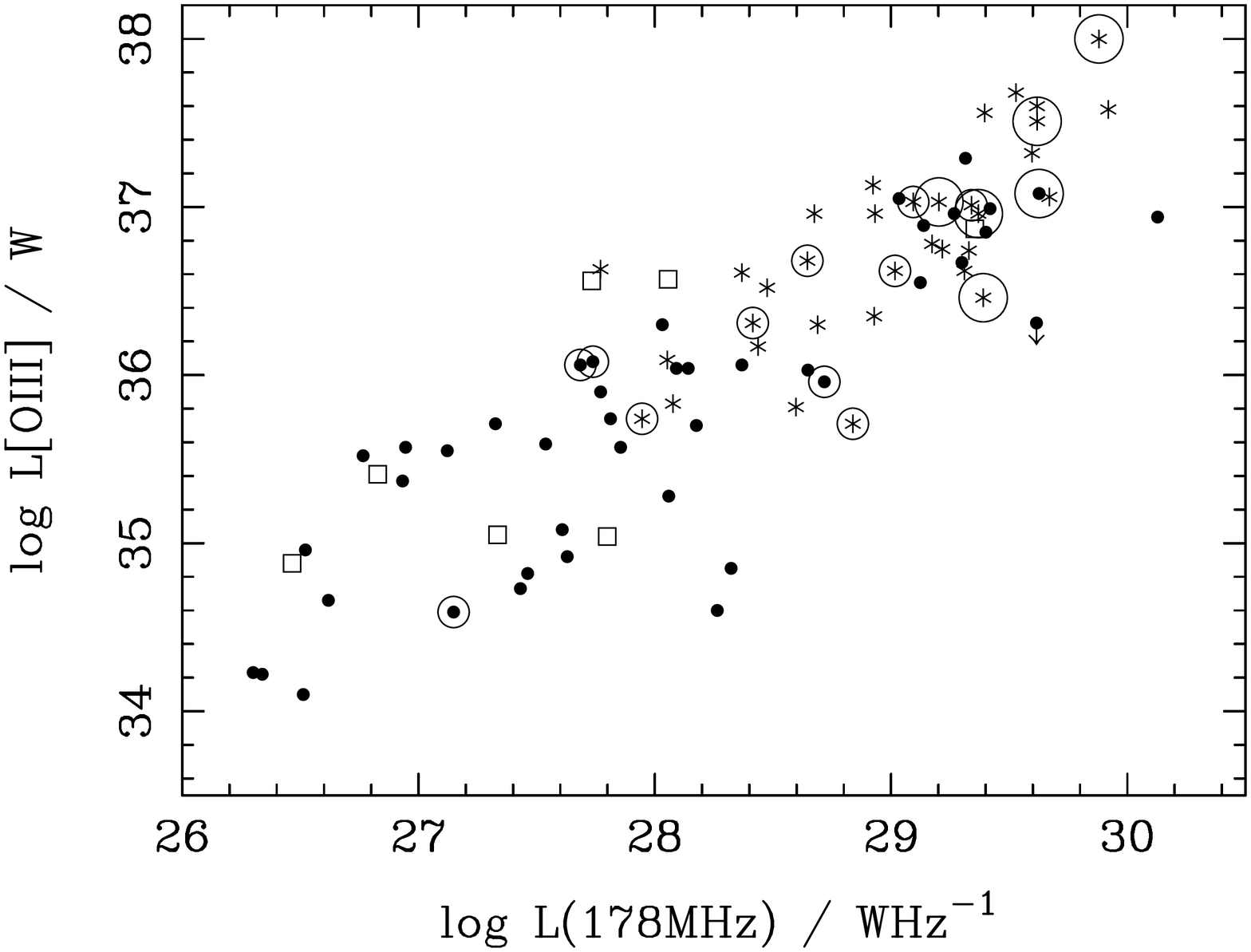,width=8.5cm,angle=-90}
\caption{\othree\ versus 178-MHz luminosity, for 3CRR sources plus OQ172
and 4C31.38. 
Quasars are denoted by asterisks, high-excitation emission-line galaxies
(HEGs) by solid circles and weak quasars (WQs) by open squares.
CSS sources are ringed
high-redshift sources from this paper. The highest-$L_{178}$
CSS source is OQ~172 - the $L_{178}$ for this source was extrapolated from the
5GHz, 408MHz and 365MHz fluxes present in the NASA Extragalactic
Database using a second-order polynomial fit.}
\label{fig:lolr}
\end{figure}

In Figure~\ref{fig:lolr} we plot \othree\ luminosity against 
178-MHz radio luminosity for the complete subset of the 3CR sources
presented by Laing, Riley \& Longair (1983, hereafter the ``3CRR sample'')
and which have published 
\othree\ fluxes. For the line strengths we take data from our observed 
CSS sample and the samples of lower redshift CSS objects from 
Gelderman \& Whittle (1994) and Jackson \& Browne (1991), 
and the larger sample of non-CSS quasars and high-excitation radio 
galaxies from Jackson \& Rawlings (1997). We have also included fluxes
from the literature (references given in Table 3 of Jackson \& Rawlings
1997 and in Table 4 of this paper) when these are available.
The Gelderman et al. (1994) sample includes some 3CRR sources but follows
the selection of Spencer et al. (1989), which consists of sources from
the Peacock \& Wall (1982) and 3CR surveys with radio structure confined
to $<$30kpc, radio spectra steeper than $\alpha=-0.5$ between 2.7 and
5~GHz, but with an additional optical magnitude cut $m_{V}<21$.

In addition to existing 3CRR data, 
Morganti et al. (1997) present observations of a low-redshift southern
sample which is selected on the same 
basis as our higher-redshift sample, by following the Fanti et al.
(1990) selection criteria, but with a redshift limit to allow
spectroscopy of \othree\ in the optical. We do not plot their data on
Fig. 4, but its inclusion does not significantly change any of the
conclusions which follow. It is also difficult to be consistent in
extrapolating the radio fluxes of their sample to 178~MHz.

We do not plot the low-excitation radio galaxies, which are
characterised by weak line emission associated with FRII radio structure
(e.g. Laing et al. 1994) as there do not appear to be any CSS LEGs. 
We only use the standard high-excitation radio galaxies (HEGs) which
show strong narrow line emission. We include as quasars the so-called
``broad-line radio galaxies'', which were referred to as ``weak quasars''
by Jackson \& Rawlings (1997) and are defined as
objects for which at least one broad line has been observed and whose
optical continuum luminosity is $<10^{23}$~W$\,$Hz$^{-1}$. 
The 178-MHz luminosities in all objects arise from extended
emission which is unaffected by relativistic beaming. A least-squares
fit to all the objects where \othree\ is detected yields the relation
log$L$(\othree ) = 13.71+0.791$\,\log L_{178}$. If we compare the scatter
about this line of \othree\ luminosities of CSS and non-CSS objects, a
Kolmogorov-Smirnov test gives no difference in the
distributions\footnote{Using the Morganti et al. (1997) sources and
extrapolating the radio fluxes of their sample down to 178~MHz gives a
difference significant 
only at the 15\% level. This modest difference is likely to be due to
the extrapolation to 178~MHz of the CSS fluxes producing an overestimate
of the $L_{\rm 178MHz}$ for these objects due to the real spectral shape
being curved and concave rather than a single power law.
Dividing the fluxes by the likely
factor of 2--3 for the overestimate removes this difference}.
We conclude that there is no firm evidence
for a significant difference between the distributions of
CSS and non-CSS sources in this diagram. 

\begin{figure}
\psfig{figure=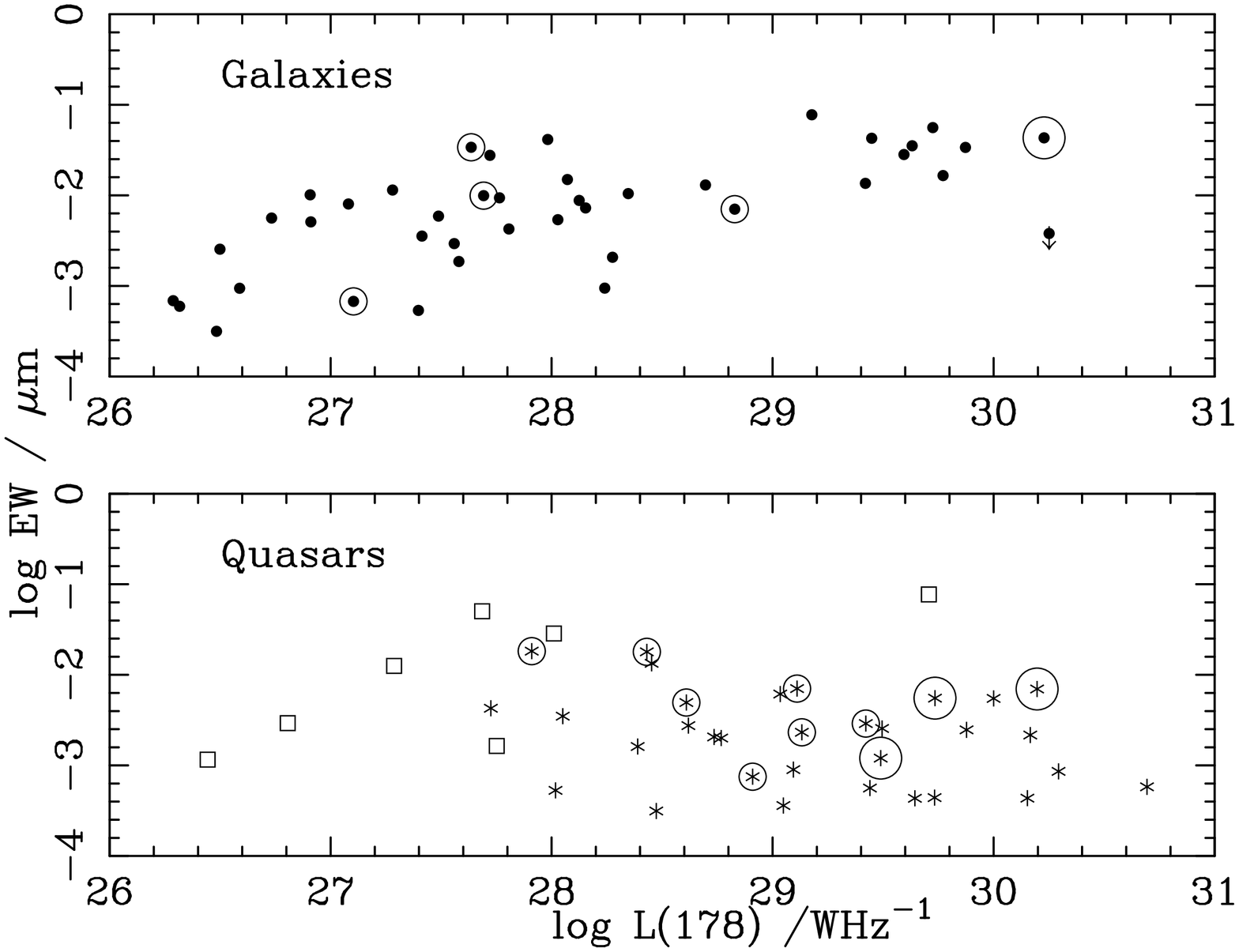,width=8.5cm,angle=-90}
\caption{\othree\ equivalent width versus 178-MHz luminosity. Symbols are as
for figure~\ref{fig:lolr}. 
OQ172 has been excluded due to the poor quality of our
spectrum which makes its optical continuum level difficult to determine.}
\label{fig:ewlr}
\end{figure}

In figure~\ref{fig:ewlr} we plot rest-frame \othree\ equivalent width 
versus 178-MHz
radio luminosity for the same sources as in figure~\ref{fig:lolr},
separating quasars and radio galaxies. A Kolmogorov-Smirnov test is
again unable to distinguish the distribution of equivalent width in either
subsample as a whole. It seems that CSS quasars have the same \othree\ 
equivalent widths and luminosities as larger radio quasars. 
If we take only the low-luminosity, low-redshift
galaxies, we cannot make a definite statement about the equivalent
widths of CSS galaxies unless we include the Morganti et al. (1997)
data; in this case the CSS galaxies have higher equivalent widths at the 0.5\%
confidence level, the result previously found by Baker \& Hunstead
(1995).

In Table ~\ref{tab:table} we present a summary of \otwo\ and \othree\ line
fluxes for the CSS subsample of 3C. This table is a supplement to that
given by Jackson \& Rawlings (1997) for the non-CSS paper of the 3CR
sample and is given in the same format.

\begin{table*}
\begin{minipage}[t]{180mm}
\caption[3C CSS objects]{3C CSS objects. Classification keys are: Q=Quasar
WQ=Weak Quasar HEG=High Excitation Galaxy (see Jackson \& Rawlings 1997)
Objects classed as HEG? are classified as N-Galaxies by SDMA. Redshifts and 
Optical Magnitudes are also from the Classification Reference; 
(R) Indicates R rather than V band magnitude.
Radio flux densities are in Jy at 178-MHz from Kellerman et al. (1969)
Line luminosities are in log$_{10}$ W.
The \othree\ luminosity refers to \othree 5007\ and does not include 
contributions from \othree 4959.

References are: 
GW94 = Gelderman \& Whittle 1994; 
HJR = This paper; 
JB91 = Jackson \& Browne 1991;
L94 = Laing et al. 1994;
MSvB = McCarthy, Spinrad \& van Breugel 1995;
SDMA = Spinrad et al. 1985;
SK96 = Stickel et al. 1996;
WRJ = Willott, Rawlings \& Jarvis 2000. 
Radio morphologies for most of these sources can be found in Spencer et
al. (1989) and Saikia et al. (1995).
}

\label{tab:table}
\begin{tabular}{|l|l|l|l|l|l|l|l|l|l|}
\hline
Object	& Class & Classification & Redshift & Optical   & Radio  & \otwo      
& \otwo     & \othree    & \othree   \\
Name    &       & Reference      &          & Magnitude & Flux   & Luminosity & 
Reference & Luminosity & Reference \\
\hline
3C43	& Q	& SDMA	& 1.47	& 20	& 11.6	&	&	&	&	\\
3C48	& Q	& SDMA	& 0.367	& 16.2	& 55.0	& 35.50	& JB91
	& 36.37	& JB91	\\
3C49	& HEG	& SDMA	& 0.621	& 21	& 10.3	&	&	&	&	\\
3C67	& WQ	& L94	& 0.310 & 18.0	& 10.0	& 35.51	& GW94	& 
36.08	& GW94	\\
3C93.1	& HEG	& SDMA	& 0.244	& 19.0	& 9.9	&	&	& 35.28
	& GW94	\\
3C99	& HEG?	& SDMA	& 0.426	& 19.1	& 10.8	&	&	&
	&	\\ 
3C119	& HEG	& SDMA	& 0.408	& 20	& 15.7	&	&	&	&	\\
3C138	& Q	& SDMA	& 0.759	& 17.9	& 22.2	& 35.82	& JB91
	& 36.71	& JB91	\\
3C147	& Q	& SDMA	& 0.545	& 16.9	& 60.5	& 36.69	& JB91
	& 37.03	& JB91	\\
3C186	& Q	& SDMA	& 1.063	& 17.6	& 14.1	&	&	& 37.03
	& HJR	\\
3C190	& Q	& SDMA	& 1.197	& 20	& 15.0	&	&	& 36.96	& 
HJR	\\
3C216	& Q	& SDMA	& 0.67	& 18.48	& 20.2	& 35.68	& JB91
	&$<$35.71& JB91\\
3C237	& HEG	& SDMA	& 0.877	& (R)21	& 20.9	&	&	&	&
	\\
3C241	& HEG	& SDMA	& 1.617	& 23.5	& 11.6	&	&	& 37.08
	& HJR	\\
3C268.3	& WQ	& L94	& 0.371	& 20.0	& 10.7	&	&	& 
35.74	& MSvB	\\
3C277.1	& Q	& SDMA	& 0.320	& 17.93	& 8.5	& 35.80	& GW94
	& 36.26	& GW94	\\
3C286	& Q	& SDMA	& 0.849	& 17.25	& 24.0	& 35.95	& GW94
	&	&	\\
3C287	& Q	& SDMA	& 1.055	& 17.67	& 16.0	&	&	&	&
	\\
3C298	& Q	& SDMA	& 1.439	& 16.79	& 47.5	&	&	&	&
	\\
3C299	& HEG	& SDMA	& 0.367	& 19.48	& 11.8	& 35.91	& MSvB
	&	&	\\
3C303.1	& HEG	& SDMA	& 0.267	& 19	& 12.4	&	&	& 36.06
	& MSvB	\\
3C305.1	& HEG	& SDMA	& 1.132	& 21.37	& 12.4	& 36.81	& 
MSvB	&	&	\\
3C309.1	& Q	& SDMA	& 0.904	& 16.78	& 22.7	&	&	&
	&	\\
3C318	& Q	& WRJ	& 1.574	& 20.3	& 12.3	&	&	& 37.51
& WRJ\\
3C343	& Q	& SDMA	& 0.988	& 20.61	& 12.4	&	&	&	&
	\\
3C343.1	& HEG	& SDMA	& 0.750	& 20.71	& 11.5	& 35.69	& 
MSvB	& 35.96	& GW94	\\
3C346	& HEG	& SDMA	& 0.161	& 17.2	& 10.9	&	&	& 34.59
	& GW94	\\
3C380	& Q	& SDMA	& 0.691	& 16.81	& 59.4	& 36.24	& GW94
	& 37.01	& GW94	\\
3C454	& Q	& SDMA	& 1.757	& 18.47	& 11.6	&	&	&	&
	\\
3C454.1	& HEG	& SK96	& 1.841	& 22.0	& 9.8	&	&	&	&
	\\ 
3C455	& Q	& SDMA	& 0.543 & 19.7	& 12.8	& 36.05	& GW94	& 
36.31	& GW94	\\
\hline
\end{tabular}
\end{minipage}
\end{table*}

\section{DISCUSSION}

\subsection{Quasar fractions among small sources}

We find possible evidence for broad H$\alpha$ in 3C~241, which has previously
been classed as a galaxy. This continues a trend seen in some other 3C
radio galaxies which have been examined by infrared spectroscopy or 
intensively with optical spectroscopy, and in which an obscured or faint
broad line region appears. Examples are 3C~318 (Willott, Rawlings \&
Jarvis 2000) which has been reclassified as a $z$=1.57 quasar, 3C~22
(Rawlings et al. 1995), 3C~67 and 3C~268.3 (Laing et al. 1994) 
in which spectroscopy revealed broad-line regions obscured by
$A_V\sim1-5$. Hill, Goodrich \& Depoy (1996) found broad Pa$\alpha$ 
in a further three radio galaxies (3C~184.1, 3C~219 and 3C~223). This
is the major reason for the removal of radio galaxies from the 3CRR 
sample with respect to earlier studies such as Saikia et al. (2001).

The significance of this can be seen by examination of Fig. 6, which
is a plot of radio power at 178~MHz against linear size of the
radio source $D$. The striking feature of this plot is that compact
sources [$D<$15~kpc for $H_0$=100~km$\,$s$^{-1}$Mpc$^{-1}$, $q_0$=1;
Fanti et al. (1990)] are predominantly (12/15) 
quasars. Including 3C~241 (marked on Fig. 6) as a galaxy, the hypothesis
that CSS and larger sources have the same quasar fraction can be
rejected at the 0.1\% level. If 3C~241 really contains a broad 
line, this increases
the power of the rejection to below 0.001\%. This result appears to be
independent of radio power effects. If we exclude objects with
$L_{178}<10^{28}$~WHz$^{-1}$, we still obtain 0.1\% significance for an
excess of quasars among small sources, and again we obtain much 
higher significance if 3C~241 is regarded as a quasar.

\begin{figure*}
\begin{tabular}{cc}
\psfig{figure=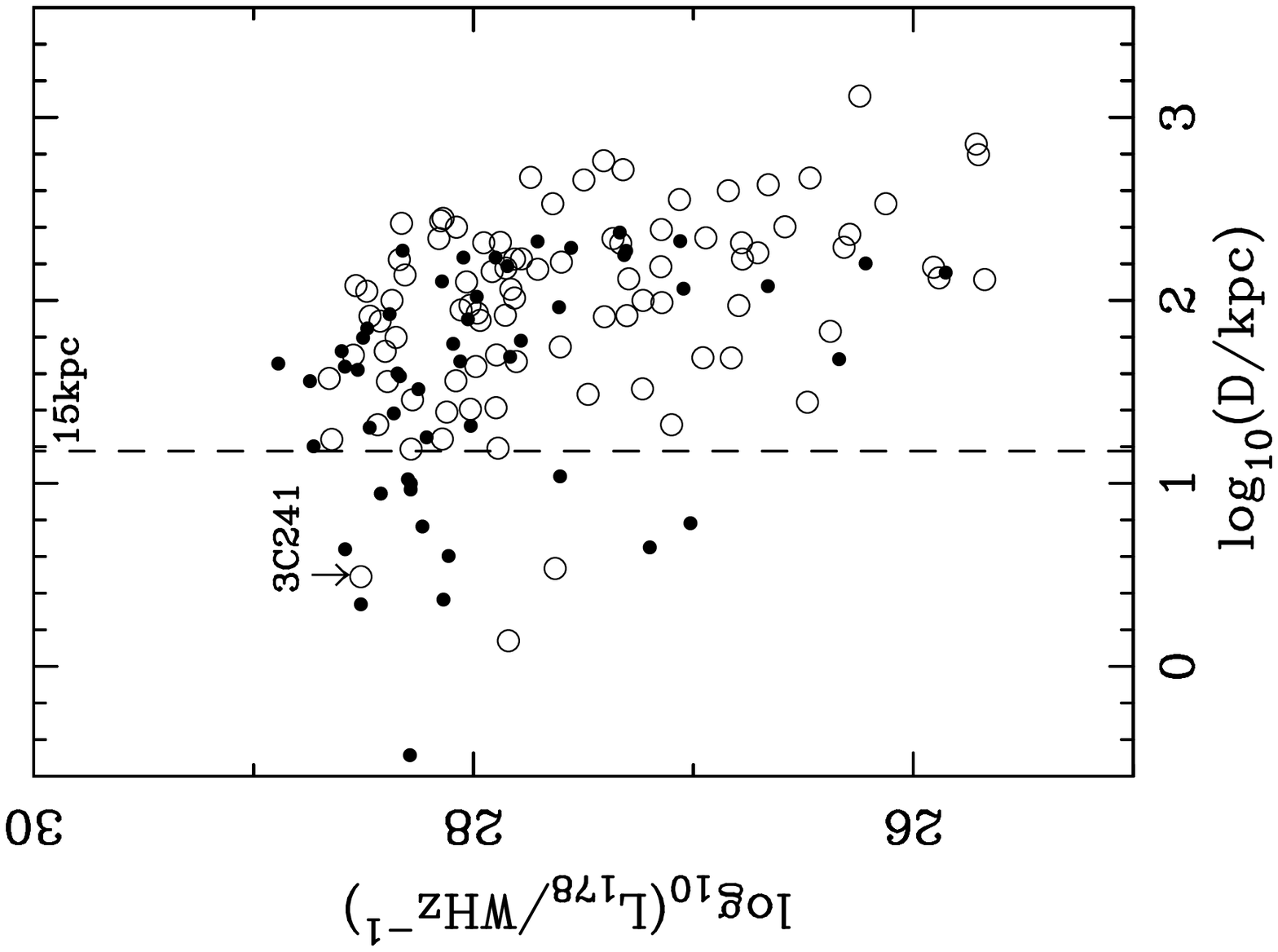,width=8cm,angle=-90}&\psfig{figure=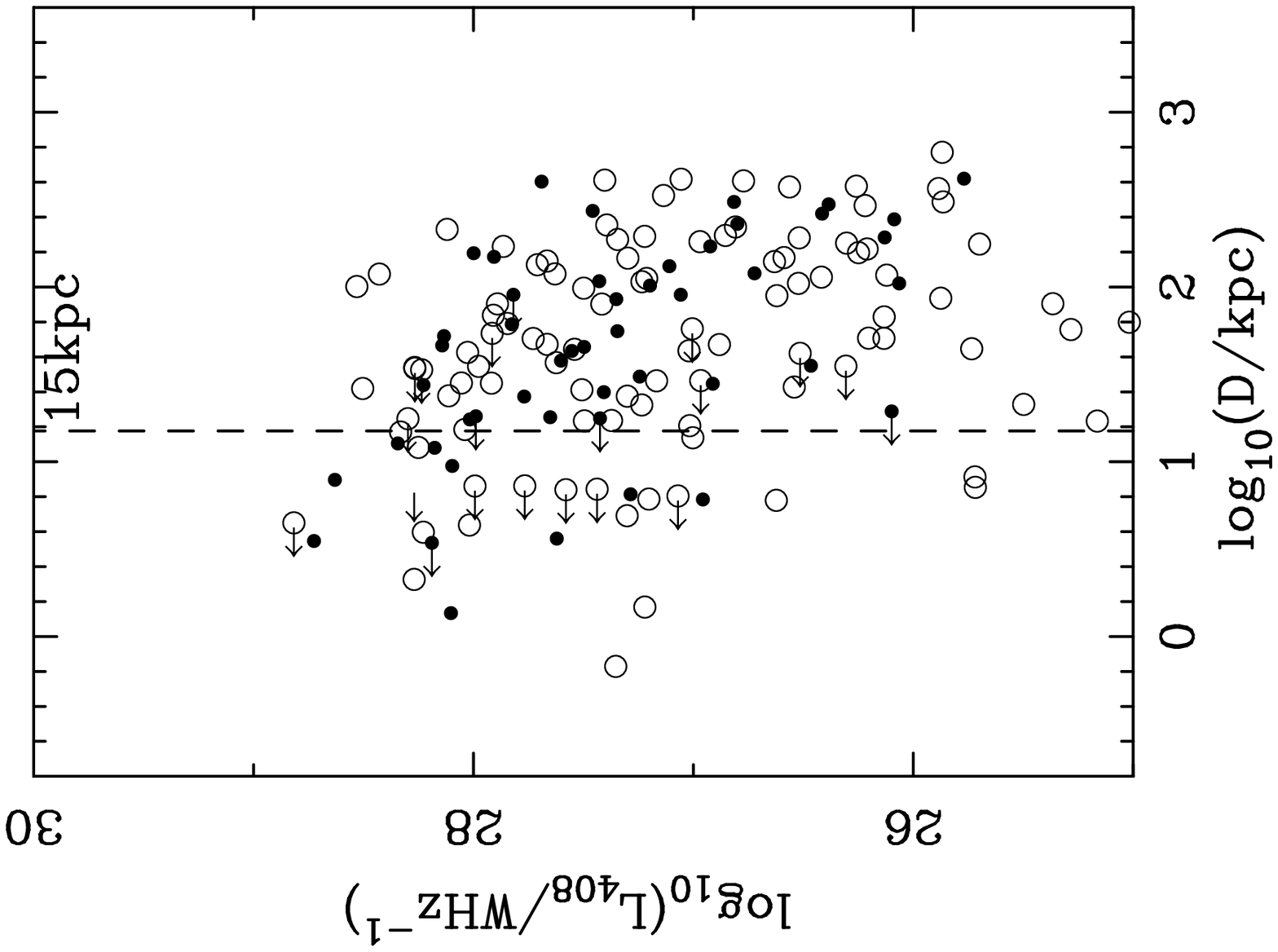,width=8cm,angle=-90}\\
\end{tabular}
\caption{
178-MHz luminosity versus projected linear size
$D$ for quasars, including WQs as defined 
by Jackson \& Rawlings (1997), (filled circles) and
radio galaxies (open circles). The 3CRR sample of Laing et al. (1983) is
plotted in the left panel and the 408-MHz sample of Best, R\"ottgering
\& Lehnert (1999) in the right panel. Powerful CSS sources are those in the 
upper-left quadrant of these plots. For these plots alone, we have used the
Fanti et al. $H_0$=100~km$\,$s$^{-1}$Mpc$^{-1}$, $q_0$=1 cosmology in
order to preserve the original $D<$15~kpc definition of a CSS source.
3C~241 is marked (see text). 3C~22, 3C~67 and 3C~268.3 have all been
considered as quasars (Laing et al. 1994, Rawlings et al. 1995).
}
\label{fig:steve}
\end{figure*}

Two of the remaining powerful CSS galaxies, 3C~43 and 3C~343.1, lack any
published modern spectra at optical wavelengths, and any near-infrared
spectra. Given the compact features visible in HST images of these
objects (de Vries et al. 1999) it is plausible that modern spectra would
reveal broad Balmer lines in these objects. We therefore find it
plausible that {\em all} the 3CR CSS sources have naked or
lightly-reddened quasar nuclei. This must be compared with a robust
quasar fraction of only $\sim$0.4 for the larger sources (Willott et al.
2000). The 3.5-$\mu$m study of powerful 3C double radio galaxies of
Simpson, Rawlings \& Lacy (1999) proves that any quasar nuclei in such
objects are hidden by an obscuring column corresponding to more than
$A_V\sim10$ magnitudes of visual extinction: any quasar nuclei in these
objects are not lightly reddened and are presumably hidden by something
more akin to an obscuring torus.

The nearest comparable sample to the 3CRR is the 408-MHz sample of
equatorial sources defined by Best, R\"ottgering \& Lehnert (1999) with
a limiting flux $S_{\rm 408MHz}>5$~Jy. This sample is virtually
spectroscopically complete and, like the 3CRR (e.g. Blundell et al. 2003,
in preparation), has radio information
that is good enough to make the angular size information unambiguous in
most cases, from VLA radio imaging by Best et al. (1999) 
and from other information (e.g.
Gower \& Hutchings 1984; Reid, Kronberg \& Perley 1999; Jeyakumar et al.
2000). It does not show the same effect. Either the 3CRR result is a
fluke, or further investigation of the 408-MHz Best et al. sample has yet to
reveal broad lines in some of the smaller sources. It may also be that
some of the Best et al. sources are somewhat younger objects with a higher
turnover frequency, which would be preferentially selected at 408 MHz,
and in which the quasar nucleus is still too deeply buried to exhibit
broad Balmer lines.

Fanti et al. (1990) conclude on statistical grounds that only a small
fraction of apparently small quasars are seen foreshortened by
projection; hence a high quasar fraction among small sources could be a
combination of a luminosity:size anticorrelation and a flux-limited
sample picking out only the brightest sources at high redshift.
Alternatively, in the popular model in which the jet-triggering
event is a galaxy-galaxy merger or interaction in which substantial
gas and dust are delivered into a region around a 
supermassive black hole, it is dynamically plausible (Willott et al.\ 2001)
that, during the first $\sim 10^{7} ~ \rm yr$ after the jet-triggering
event, this material is distributed as a relatively extended mist of
translucent gas/dust clouds. These clouds might be 
capable of lightly obscuring the nucleus, but incapable of
blocking the light completely so that 
one would expect all CSS sources to have visible quasar nuclei,
albeit with a small amount of reddening by the extended dust.
As the central region dynamically relaxes, and the radio source expands, 
the standard optically-thick obscuring torus can form, and observations of the
quasar nuclei of these larger double sources will then be in accord with the
standard unified scheme for radio sources.

\subsection{Luminosity and equivalent width of \othree }

We have found that \othree\ line luminosities of objects at high 
redshift for CSS are not significantly different from those of larger
radio sources.

This confirms a number of previous results at low
redshift by Baker and Hunstead (1995) and also Morganti et al. (1997);
the latter authors however suggest tentative evidence for slightly 
lower line luminosity for CSS. It also agrees with the same result 
for Ly$\alpha$ by Jarvis et al. (2001). There are several implications
of this result. The direct inference is that the underlying quasar 
luminosity, which is responsible for emission-line excitation, either by
photoionization (e.g. Davidson \& Netzer 1979) or by driving shocks into
the surrounding gas (Sutherland, Dopita \& Bicknell 1993),
remains roughly constant as quasars age. It also makes the
``frustration'' scenario, in which CSSs are viewed as radio sources
which have been prevented from escaping from the galactic medium (van
Breugel et al. 1984) difficult to sustain as in this case increased line
emission would be expected corresponding to the increased level of
interaction between the radio jet and its surroundings 
(Morganti et al. 1997).

We also find that there is no tendency for CSS galaxies or quasars to
have different equivalent widths at high redshift from their larger
cousins. In Fig. 7 we plot the \othree\ equivalent width against the
degree of radio core-dominance $R$, defined as the emitted-frame ratio
of core and extended radio flux, in order to separate out the tendency
of quasars to have higher values of $R$.

This is a more interesting result, because at low redshift and in a
larger sample, Baker \&
Hunstead (1995) find a higher equivalent width for CSSs and suggest that
this is due to suppression of the optical continuum in CSS by reddening.
We can perhaps understand why there is a difference between the \othree\
equivalent widths of CSS and non-CSS sources at low redshift, but may not
be at high redshift. Blundell \& Rawlings (1999) have pointed out that the
highest-redshift objects in any flux-limited sample, like 3C, will
inevitably be young, short radio sources. Hence there is much less
difference between the CSS sources and their larger counterparts in
terms of both time since the jet-triggering event and the radii to which
radio-source shocks could have propagated.

\begin{figure}
\psfig{figure=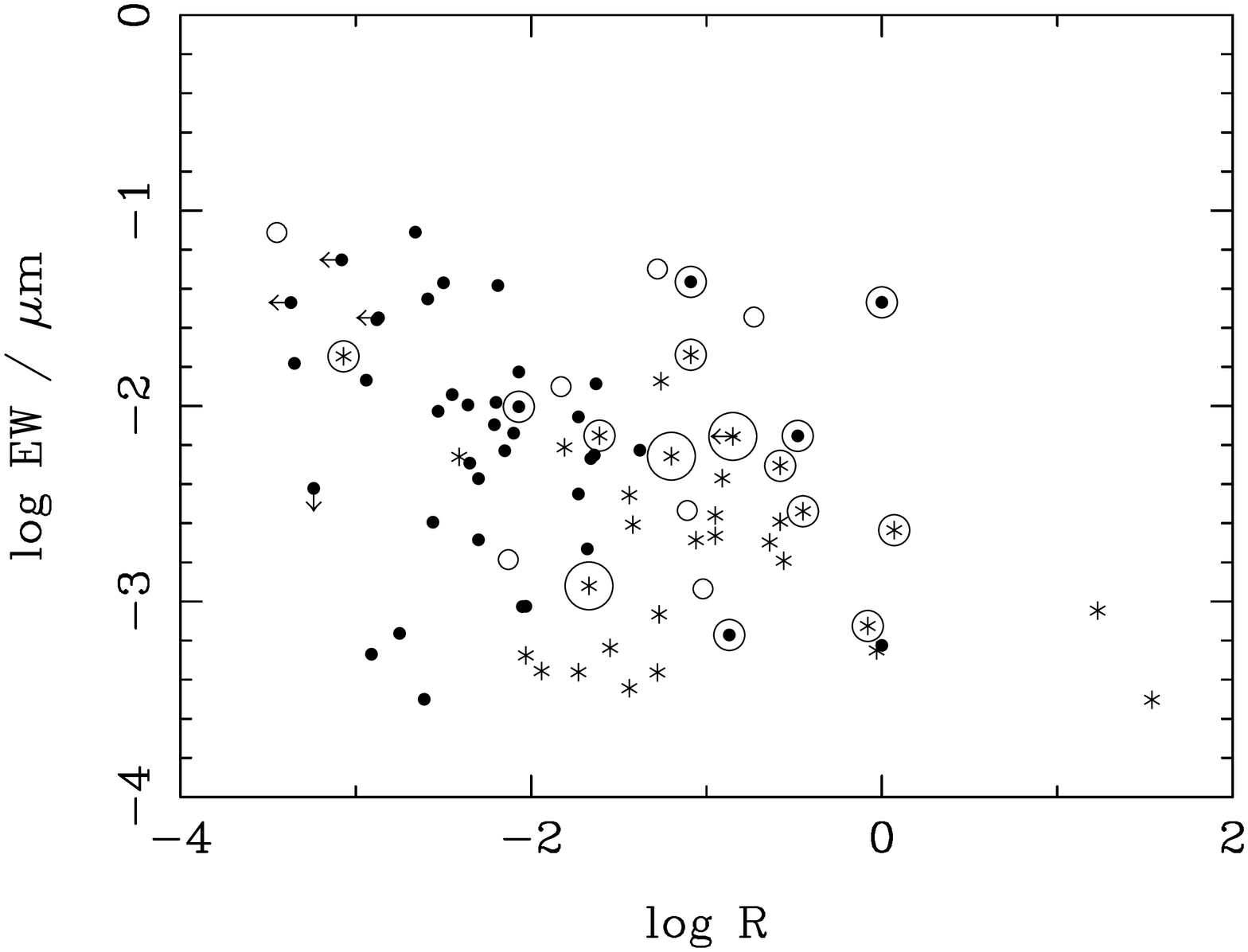,width=8.5cm,angle=-90}
\caption{\othree\ equivalent width versus $R$. Symbols are as
for figure~\ref{fig:lolr}, $R$ values have all been converted to 8GHz
emitted, assuming $\alpha=0$ for the
core and $\alpha=1$ for the extended emission.}
\label{fig:ewr}
\end{figure}

\section{CONCLUSIONS}

We have presented new near-infrared spectra of nine radio-loud, compact
steep-spectrum objects, six radio galaxies and two quasars. The narrow-line
\othree\ luminosities in these mostly high-redshift sources are
comparable to those in larger sources, suggesting that whether the
excitation mechanism for the lines is mechanical energy input via shocks
or input of high-energy photons from the central engine, it remains of 
roughly constant power
as the radio sources expand and age. The H$\beta$ line in one radio
galaxy may be broad; even without this object, the quasar fraction of
small sources in the 3CRR complete sample is now very high.

\medskip
\noindent {\bf Acknowledgements}
\medskip

The United Kingdom Infrared Telescope is operated by the Joint Astronomy 
Centre on behalf of the U.K. Particle Physics and Astronomy Research Council. 
We thank the support staff at this telescope. This research has made use
of the NASA/IPAC Extragalactic Database (NED) which is operated by the 
Jet Propulsion Laboratory, California Institute of Technology, under 
contract with the National Aeronautics and Space Administration. We
thank an anonymous referee for useful comments on the paper. 

\medskip
\noindent {\bf References}
\medskip

\noindent Alexander P., 2000, MNRAS, 319, 8

\noindent Axon D.J., Capetti A., Fanti R., Morganti R., Robinson A.,
Spencer R.E., 2000, AJ, 120, 2284

\noindent Baker A.C., Carswell R.F., Bailey J.A., Espey B.R., Smith
M.G., Ward M.J., 1994, MNRAS, 270, 575

\noindent Baker J.C., Hunstead R.W., 1995, ApJ, 452, L95

\noindent Baker J.C., Hunstead R.W., 1996, in Proc. 2nd Workshop on 
GPS and CSS sources, Leiden, ed. I. Snellen

\noindent Baker J.C., Hunstead R.W., Kapahi V.K., Subrahmanya C.R.,
1999, ApJS, 122, 29

\noindent Barthel P.D., Tytler D.R., Thomson B., 1990, A\&AS, 82, 339

\noindent Best P., R\"ottgering H.J.A., Longair M.S., 2000, MNRAS, 311, 23

\noindent Best P., R\"ottgering H.J.A., Lehnert M.D., 1999, MNRAS, 310, 223

\noindent Bicknell G.V., Dopita M.A., O'Dea C.P., ApJ, 485, 112

\noindent Blundell K.M., Rawlings S., 1999, Nature, 399, 330

\noindent van Breugel W., Miley G., Heckman T., 1984, AJ, 89, 5

\noindent Carvalho J.C. 1985, MNRAS, 215, 463

\noindent Davidson K., Netzer H., 1979, Rev. Mod. Phys., 51, 715

\noindent Eales S.A., Rawlings S. 1993, ApJ, 411, 67

\noindent Fanti C., Fanti R., Schilizzi R.T., Spencer R.E., van Breugel W.J.M.,
1986, A\&A, 170, 10

\noindent Fanti C., et al., 1989, A\&A, 217, 44

\noindent Fanti R., Fanti C., Schilizzi R.T., Spencer R.E., Rendong N., 1990,
A\&A, 231, 333

\noindent Fanti C., et al., 2000, A\&A, 358, 499

\noindent Gelderman R., Whittle M., 1994, ApJS, 91, 491

\noindent Gower A., Hutchings J., 1984, AJ, 89, 1658

\noindent Heckman T.M., 1980, A\&A, 88, 311

\noindent Hes R., Barthel P.D., Fosbury R.A.E., 1996, A\&A, 313, 423

\noindent Hill G.J., Goodrich R.W., Depoy D.L., 1996, ApJ, 462, 163

\noindent Inskip K.J., Best P.N., Rawlings S., Longair M.S., Cotter G.,
R\"ottgering H.J.A., Eales S., 2002, MNRAS, 337, 1381

\noindent Jackson N., et al., 1990, Nature, 343, 43

\noindent Jackson N., Browne I.W.A., 1991, MNRAS, 250, 422

\noindent Jackson N., Eracleous M., 1995, MNRAS, 276, 1409

\noindent Jackson N., Rawlings S., 1997, MNRAS, 286, 241

\noindent Jarvis M., et al., 2001, MNRAS, 326, 1563

\noindent Jenkins C.J., Pooley G.G., Riley J.M., 1977, MemRAS, 84, 61

\noindent Jeyakumar S., Saikia D.j., Pramesh Rao A., Balasubhramanian
V., 2000, A\&A, 362, 27

\noindent Kapahi V.K., Athreya R.M., Subrahmanya C.R., Baker J.C.,
Hunstead R.W., McCarthy P.J., van Breugel W., 1998, ApJS, 118, 327

\noindent Kellerman K.I., Pauliny-Toth I.I.K., Williams P.J.S., 1969,
ApJ, 157, 1

\noindent K\"uhr H., Witzel A., Pauliny-Toth I.I.K., Nauber U., 1981,
A\&AS, 45, 367

\noindent Kuraszkiewicz J., Wilkes B.J., Brandt W.N., Vestergaard M.,
2000, ApJ, 542, 631

\noindent Laing R.A., Jenkins C.R., Wall J.V., Unger S.W., 1994, in
Proc. First Stromlo Symposium ``The physics of active galaxies'', ASP
Conf. Ser. vol. 54, eds. Bicknell G.V. et al., p. 201

\noindent Laing R.A. Riley J., Longair M.S., 1983, MNRAS, 204, 151

\noindent Lawrence A., 1991, MNRAS, 252, 586

\noindent McCarthy P., Spinrad H, van Breugel W.J.M., 1995, ApJS, 99, 27

\noindent Miley G.K., Miller J.S., 1979, ApJ, 228, L55

\noindent Morganti R., Tadhunter C.N., Dickson R., Shaw M., 1997, A\&A, 326, 130

\noindent Morton D.C.; Peterson B.A., Chen J-S, Wright A.E.,
Jauncey, D.L., 1989, MNRAS, 241, 595

\noindent Mountain C.M., Robertson D.J., Lee T.J., Wade R., 1990, Proc
SPIE

\noindent O'Dea C.P., Baum S.A., Stanghellini C., 1991, ApJ, 380, 66

\noindent O'Dea C.P., Baum S.A., 1997, AJ, 113, 148

\noindent O'Dea C.P., 1998, PASP, 110, 493

\noindent O'Dea C.P., et al. 2002, AJ, 123, 2333

\noindent Orr M.J.L., Browne I.W.A., 1982, MNRAS, 200, 1067

\noindent Owsianik I., Conway J.E., 1998, A\&A, 337, 69

\noindent Peacock J.A., Wall J.V., 1982, MNRAS, 198, 843

\noindent Phillips R.B., Mutel R.L., 1982,  A\&A, 106, 21

\noindent Pilkington J.D.H., Scott P.F., 1965, MemRAS, 69, 183

\noindent Rawlings S., Lacy M., Sivia D.S., Eales S.A., 1995, MNRAS,
274, 428

\noindent Reid R.I., Kronberg P.P., Perley R.A., 1999, ApJS, 124, 285

\noindent di Serego Alighieri S., Cimatti A., Fosbury R.A.E., Hes R., 
1997, A\&A, 328, 510

\noindent Saikia D.J., Jeyakumar S., Wiita P.J., Sanghera H.S., Spencer 
R.E., 1995, MNRAS, 276, 1215

\noindent Saikia D.J., Jeyakumar S., Salter C.J., Thomasson P., Spencer
R.E., Mantovani F., 2001, MNRAS, 321, 37

\noindent Simpson C. 1998, MNRAS, 297, L39

\noindent Simpson C., Rawlings S., Lacy M., 1999, MNRAS, 306, 828

\noindent Snellen I.A.G., Schilizzi R.T., Miley G.K., de Bruyn A.G., 
Bremer M.N., R\"ottgering H.J.A., 2000, MNRAS, 319, 445

\noindent Spencer R.E., McDowell J.C., Charlesworth M., Fanti C., Parma
P., Peacock J.A., 1989, MNRAS, 240, 657

\noindent Spinrad H., Djorgovski G., Marr J., Aguilar L., 1985, PASP,
97, 932

\noindent Stickel M., K\"uhr H., 1996, A\&AS, 115, 1

\noindent Sutherland R.S., Bicknell G.V., Dopita M.A., 1993, ApJ, 414, 510

\noindent Udomprasert P.S., Taylor G.B., Pearson T.J., Goberts D.H.,
1997, ApJ, 483, L9

\noindent de Vries W.H., et al. 1998, ApJ, 503, 138

\noindent de Vries W.H., O'Dea C.P., Baum S.A. Barthel P.D., 1999, ApJ, 526, 27

\noindent Willott C., Rawlings S., Blundell K., Lacy M., 2000, MNRAS,
316, 449

\noindent Willott C., Rawlings S., Jarvis M., 2000, MNRAS, 313, 237

\noindent Willott C., Rawlings S., Blundell K.M., Lacy M., Eales S.A.,
2001, MNRAS, 322, 536

\noindent Wills B., Browne I.W.A., 1986, ApJ, 302, 56

\label{lastpage}
\end{document}